# AN ANALYSIS ON IMPROVEMENT OF X-RAY DIFFRACTOMETER RESULTS BY CONTROLLING AND CALIBRATION OF PARAMETERS

**Hamidreza Moradi[1] , Fatemeh Mehradnia[2]\***

[1] Department of Mechanical Engineering and Engineering Science, The University of North Carolina at Charlotte, Charlotte, North Carolina, USA
[2] Center for Translational Medicine, Department of Biomedical and Pharmaceutical Sciences, University of Montana, USA

**Abstract:**
The X-ray diffractometer in the laboratory is a crucial instrument for analyzing materials in science. It can be used on almost any crystal material, and if the machine parameters are appropriately controlled, it can offer a lot of information about the sample's characteristics. Nevertheless, the data obtained from these machines are complicated by an aberration function that can be resolved through calibration. In this study, a powder comprising of Barium Sulfate ($BaSO_4$), Zinc Oxide (ZnO) and Aluminum (Al) was used as the first sample and a single crystal sample comprised of Gallium Nitride (GaN) and Aluminum Oxide ($Al_2O_3$). The required calibration parameters of the X-ray diffractometer namely: Straight Beam Alignment, Beam Cut Alignment and Sample Tilt Alignment for two samples were analyzed and carried out. Using the results of the X-ray spectrum, important parameters such as corresponding planes for peak positions, d-spacing of planes, intensities, smallest crystallite sizes and lattice parameters, and a comparison with the reference data were all carried out. As another result, the out-of-plane alignment and Full-Width-at Half-Maximum (FWHM) value for GaN could be determined using the rocking curve.

## 1. INTRODUCTION

The X-ray powder diffractometer, a versatile analytical instrument employed in various scientific disciplines, including material science [1–4], geology [5], and biomedical device analysis [6–8], plays a crucial role in characterizing crystalline materials. It can collect a continuous suite of hkl reflections with a single scan in θ-2θ angle space and is suitable for small crystallites ranging from 5 μm to 30 μm. With advanced data analysis methods, it can provide both qualitative and quantitative information [9,10]. Modern commercial instruments may have additional features like focusing mirror optics and interchangeable experimental configurations.

On the one hand, conducting numerical analysis requires statistical analysis [11–15] and experimental design, susceptible to errors and barriers such as uncertainty [16]. On the other hand, every experimental work needs a good calibration and inspection [17–20]. The divergent beam X-ray diffractometer can illuminate many crystallites, resulting in a strong diffraction signal from a representative portion of the sample. However, the para-focusing optics of laboratory diffractometers produce complex profile shapes that do not necessarily reflect the proper spacing of hkl planes. Modeling these aberrations requires further analysis, but there are also instrumental effects that are not well understood and can lead to confounding results when instruments are set up incorrectly. Therefore, the use of SRMs to calibrate instrument performance is preferred to avoid these situations.

*CONTACT: F. Mehradnia, e-mail: fatemeh.mehradnia@mso.umt.edu

$BaSO_4$ powders are commonly used in coatings for the internal surfaces of spectrophotometers integrating spheres, construction materials, and household paints. These coatings are characterized by their diffuse reflection spectrum and integral absorption coefficient of solar radiation [21,22]. $BaSO_4$ powders have a high reflectance in the UV, visible, and near IR regions of the spectrum [23–25], making them ideal for reflective coatings in spectrophotometers. The approximate band gap of $BaSO_4$ is 6 eV, with the main absorption edge at around 200 nm [26]. In contrast, coatings based on ZnO and $TiO_2$ powders exhibit significantly larger solar absorptivity values, ranging from 0.15 to 0.19 for ZnO pigment-based coatings and 0.17 to 0.2 for $TiO_2$ pigment-based coatings [27]. The coefficient $α_s$ of white pigments in ZnO used in coatings typically ranges from 0.15 to 0.2 [28,29], while titanium dioxide powder-based coatings have a coefficient that can reach up to approximately 0.28 [29].

The absorption coefficient $α_s$ can be reduced by increasing the reflection coefficient $ρ$ of the powders in longer wavelength regions than the main absorption edge of the solar spectrum. The size of powder particles and impurities in the powder pigment also affect $ρ$ in different spectrum regions.

For a powder to have a high $ρ$ value across a broad spectrum, such as the solar one (0.2–3.0 μm), it must be very pure and have particles that closely match this spectrum [30]. Adding nanoparticles is one way to alter how big the powder particles are and prevent them from losing their optical qualities when exposed to space radiation [31,32]. This helps achieve the maximum increase in their stability under radiation [33].

Some nanopowders that can modify pigments are ZnO, $ZrO_2$, Al, etc. [34]. However, the optical qualities of $BaSO_4$ powders and other substances that are modified by these nanopowders may change differently. The modification may only improve their reflectivity a little. That is why such research is scientifically and practically relevant. This study focuses on the first sample, which consists of $BaSO_4$, ZnO and Al.

Some insulators like $SiO_2$, $SiN_x$, $Al_2O_3$, and $HfO_2$ can help reduce gate current and protect the surface of GaN metal insulator semiconductor high electron mobility transistors (MISHEMTs) [35–37]. $Al_2O_3$ has some advantages, such as a big band gap, breakdown electric field, and simple deposition. The dielectric/GaN interface is crucial for how well GaN-based MISHEMTs perform for mm-wave and power switching uses.

The main desire in this work is to determine sources of measurement error and the procedures to calibrate the laboratory X-ray diffraction instrument properly. In this regard, first, the required alignments for the machine were carried out: Straight Beam Alignment, Beam Cut Alignment and Sample Tilt Alignment for the diffractometer. After conducting the X-ray diffraction, we could match the materials used in both samples, the powder comprised of $BaSO_4$, ZnO and Al and the single crystal sample comprised of GaN and $Al_2O_3$. The corresponding planes for peak positions, intensities, comparison with the reference data, the smallest crystallite sizes and lattice parameters were all carried out using the results of the X-ray spectrum. The rocking curve of the GaN also helped find the out-of-plane alignment and Full-Width-at Half-Maximum (FWHM) value. Notably, this work significantly contributes to the field by providing comprehensive insights and practical guidance on the calibration and alignment of X-ray diffractometers, facilitating accurate material analysis and advancing scientific research quality, making it a valuable addition to the academic literature.

## 2. MATERIALS AND METHODS

In this experiment, using X-ray diffraction, a Panalytical X'Pert PRO X-ray diffractometer matches the crystal structure and phases of two samples (powder and single crystal) using X-ray diffraction. In the first step, a radiation measurement will be conducted using a Geiger counter to ensure the safety of the experiment. A Cu radiation with $Kα_1$ wave length of 1.5406Å was used in this experiment. The voltage and current of the generator were set to be 45 KV and 40 mA, respectively.

**2.1 Loading sample and beam optics hardware**

For the first sample, which is comprised of $BaSO_4$, ZnO and Al, at the incident side of the experiment, a 1/32° divergence slit to control the intensity of the diffracted beam and a Cu (0.2 mm)/Ni (0.02 mm) attenuator to reduce the incident beam intensity by a specific factor, are used. The smallest possible divergence slit is used to prevent any deviation in the X-ray beam at low incident angles. At the receiving side, a 0.27° receiving slit and a 0.04 rad soller slit for collimation of diffracted beam and enhancing the resolution of the scans, are used. Notably, Ni in the attenuator performs as a filter for the Kβ line. It features the maximum possible Kα line while reducing the intensity of the Kβ line. On the other hand, Cu does not absorb any radiation, yet it scatters it.

For the other sample (Fig.1), which is comprised of GaN and $Al_2O_3$, first, a 1/2° divergence slit with Ni 0.02 filter is used. Type 0.27° and 0.04 rad soller slit collimators are used on the receiving side of the machine. Besides mentioned, an automatic 0.125 mm Ni beam attenuator helps in automatic insertion/removal of the attenuator.

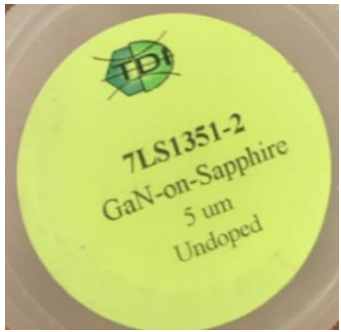

**Fig. 1.** Thin film sample

## 2.1 Beam alignment for a powder sample

*Straight beam alignment*

In this step, the 2θ angle is aligned at the 0 position of the straight beam. An I-2θ plot is derived to fulfill the alignment (Fig.2).

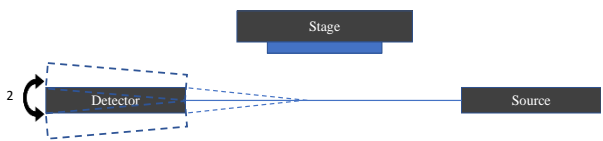

**Fig. 2.** Straight Beam Alignment

*Beam Cut Alignment*

This process, also known as Z alignment, aligns the sample's surface to the diffractometer's center (Fig.3). In this Step, an I-Z plot is generated.

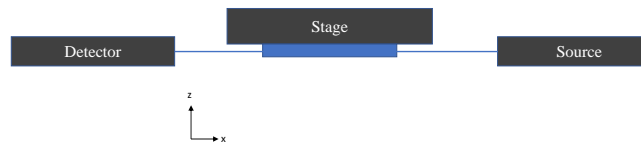

**Fig. 3.** Beam Cut Alignment

*Sample Tilt Alignment*

This procedure, which is also known as ω alignment, is performed to make the sample surface parallel to the straight beam as depicted in Fig. 4. Notably, in our case ω= θ.

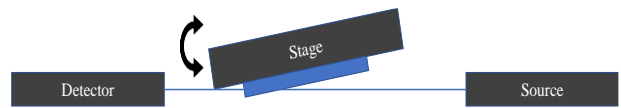

**Fig. 4.** Tilt Alignment

After the alignment, a full θ-2θ scan in the range of (20°-80°) and with 0.02 scan step size will be conducted to determine the peak positions and relative intensities of X-ray diffraction.

## 3. RESULTS AND DISCUSSION

In this section, the X-ray spectrum of both samples will be illustrated and indexed using reference data from the International Center for Diffraction Data (ICDD). In the next step, a comparison between experimental and ICDD data for intensities has been illustrated, as well as plane spacing. Besides, the smallest crystallite size of each component is calculated.

### 3.1 Powder sample

Fig. 5 shows θ-2θ full scan spectrum of X-ray diffraction as well as the index of each plane based on ICDD data for all components namely $BaSO_4$, ZnO and Al [38].

Table 1 shows the angle and intensity of peak positions in both experimental data and the ICDD file. Also, the d-spacing of planes for each material component can be seen. From Table 1 we can see that the 2θ Difference between experimental and reference data is almost constant for each material component at different angles. We see two changes for $BaSO_4$ from the reference at 1-2 and 4-5 peaks, and for Al in 1-2 peak. All the discrepancies in the spectrum could be because of thermal, residual strain, preferred orientation, and production factors [39]. As shown in Table 1 there are fewer peaks for Al and ZnO which could be due to a lower content of these components in the powder sample.

Based on the ICDD files, we know that the crystal structure for BaSO$_4$, ZnO and Al are orthorhombic, hexagonal and cubic, respectively. Therefore, we can find the corresponding crystal parameters illustrated in Table 2.

Table 3 illustrates the smallest crystallite size for the powder sample using the Scherrer equation [40] as Eq. (1):

$$t = \frac{0.9\lambda}{B\cos\theta_B} \quad (1)$$

For ZnO, BaSO$_4$ and Al, the highest intensities were found at 2θ angles of 36.23°, 42.57° and 44.59°, respectively. Therefore, these peaks were used to determine the crystallite sizes in Table 3.

**Table 1.** Powder sample crystal planes comparison with the ICDD data

| Experimental 2ϑ (°) | Real Intensity | Relative Intensity | ICDD 2ϑ (°) | ICDD Relative Intensity | 2ϑ Difference | d (Å)=λ/2sinθ | Material | hkl plane |
|---|---|---|---|---|---|---|---|---|
| 22.79 | 71 | 31.56 | 22.81 | 50 | 0.03 | 3.89 | BaSO$_4$ | 111 |
| 24.86 | 72 | 32 | 24.89 | 30 | 0.03 | 3.57 | BaSO$_4$ | 200 |
| 25.81 | 136 | 60.44 | 25.86 | 100 | 0.05 | 3.44 | BaSO$_4$ | 021 |
| 26.81 | 178 | 79.11 | 26.86 | 70 | 0.05 | 3.32 | BaSO$_4$ | 210 |
| 28.71 | 168 | 74.67 | 28.77 | 95 | 0.06 | 3.10 | BaSO$_4$ | 121 |
| 31.55 | 110 | 48.89 | 31.55 | 50 | 0 | 2.83 | BaSO$_4$ | 211 |
| 31.73 | 129 | 57.33 | 31.80 | 57 | 0.07 | 2.81 | ZnO | 100 |
| 32.77 | 58 | 25.78 | 32.82 | 45 | 0.05 | 2.73 | BaSO$_4$ | 002 |
| 34.41 | 78 | 34.67 | 34.45 | 44 | 0.04 | 2.60 | ZnO | 002 |
| 36.23 | 225 | 100 | 36.28 | 100 | 0.05 | 2.47 | ZnO | 101 |
| 38.31 | 37 | 16.44 | 38.50 | 100 | 0.19 | 2.34 | Al | 111 |
| 40.79 | 38 | 16.89 | 40.81 | 25 | 0.02 | 2.21 | BaSO$_4$ | 122 |
| 42.57 | 183 | 81.33 | 42.63 | 80 | 0.06 | 2.12 | BaSO$_4$ | 140 |
| 42.91 | 140 | 62.22 | 42.95 | 75 | 0.04 | 2.10 | BaSO$_4$ | 212 |
| 43.99 | 49 | 21.78 | 44.02 | 19 | 0.03 | 2.05 | BaSO$_4$ | 041 |
| 44.59 | 140 | 62.22 | 44.78 | 47 | 0.19 | 2.03 | Al | 200 |
| 47.53 | 45 | 20 | 47.58 | 23 | 0.05 | 1.91 | ZnO | 102 |
| 48.99 | 51 | 22.67 | 49.04 | 18 | 0.05 | 1.85 | BaSO$_4$ | 330 |
| 56.61 | 74 | 32.89 | 56.65 | 32 | 0.05 | 1.62 | ZnO | 110 |
| 62.83 | 61 | 27.11 | 62.92 | 29 | 0.09 | 1.47 | ZnO | 103 |
| 65.07 | 53 | 23.56 | 65.19 | 22 | 0.12 | 1.43 | Al | 220 |
| 67.91 | 55 | 24.44 | 68.02 | 23 | 0.11 | 1.37 | ZnO | 112 |
| 78.11 | 66 | 29.33 | 78.30 | 24 | 0.19 | 1.22 | Al | 311 |

**Table 2.** Lattice Parameters for a powder diffraction

| Material | Formula | a(Å) | b(Å) | c(Å) |
|---|---|---|---|---|
| ZnO | $\frac{1}{d^2} = \frac{4}{3}\left(\frac{h^2 + hk + l^2}{a^2}\right) + \frac{l^2}{c^2}$ | 3.24 | 3.24 | 5.2 |
| BaSO$_4$ | $\frac{1}{d^2} = \frac{h^2}{a^2} + \frac{k^2}{b^2} + \frac{l^2}{c^2}$ | 7.08 | 8.77 | 5.46 |
| Al | $d_{hkl} = \frac{a}{\sqrt{h^2 + k^2 + l^2}}$ | 4.05 | 4.05 | 4.05 |

**Table 3.** Smallest crystallite size of a powder sample

| Material | hkl plane | Relative Intensity | θ (°) | B (Rad)×10⁻³ | t (nm) |
|---|---|---|---|---|---|
| ZnO | 101 | 100 | 18.12 | 4.54 | 32.13 |
| BaSO$_4$ | 140 | 81.33 | 21.29 | 4.19 | 35.51 |
| Al | 200 | 62.22 | 22.30 | 4.89 | 30.65 |

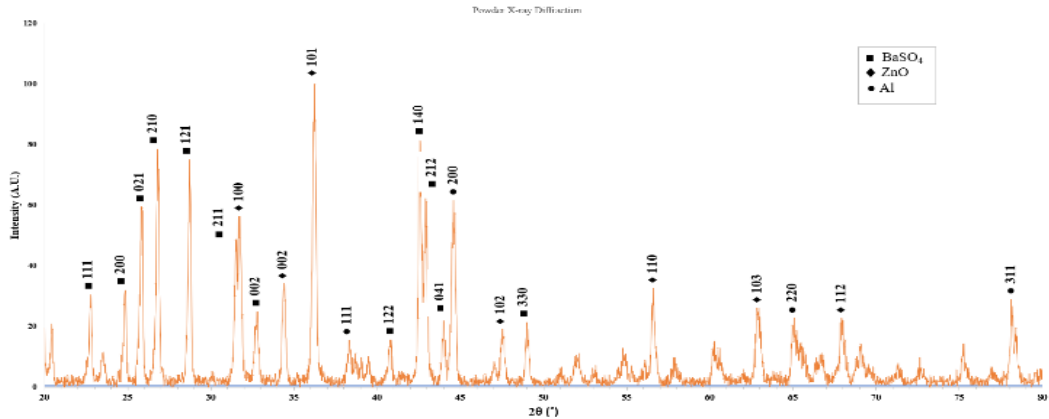

**Fig. 5.** X-ray spectrum for a powder sample

### 3.2 Thin film sample

Fig. 6 shows θ-2θ full scan spectrum of X-ray diffraction and the index of each plane based on the ICDD data for a single crystal sample with GaN and Al$_2$O$_3$ [38] components. It can be seen that just a family of planes that are parallel to (100) are illustrated. It is because of the preferred orientation effect. All other planes not parallel to (100) will be canceled from the plot.

Table 4 shows the angle and intensity of peak positions in both experimental data and the ICDD file for a single crystal sample. The d-spacing of planes for each material component can be seen, too. The results show a high difference between the film intensity (GaN) and substrate (Al$_2$O$_3$) which can be due to the beam bisection alignment method used while performing the experiment.

For GaN, Al$_2$O$_3$, the highest intensities were found at 2θ angles of 34.51° and 41.61°, respectively. Therefore, these peaks were used to determine the crystallite sizes in Table 5.

Based on the ICDD files, we know that the crystal structures for GaN and Al$_2$O$_3$ are hexagonal and Rhombohedral, respectively. Therefore, we can find the corresponding crystal parameters illustrated in Table 6.

In Fig. 7 the ω-scan for GaN, which is also called a rocking curve for the sample, can be seen. We can also see the peak center and Full-Width-at Half-Maximum (FWHM) inside the plot.

In the Fig. 7, the peak center is at 17.232°; however the data shows that the peak center is at ½ × 34.59 = 17.295. Therefore, there is 0.063° which can be because of several reasons, such as residual stress in the film. A low amount of FWHM shows that the crystallinity of the sample is good.

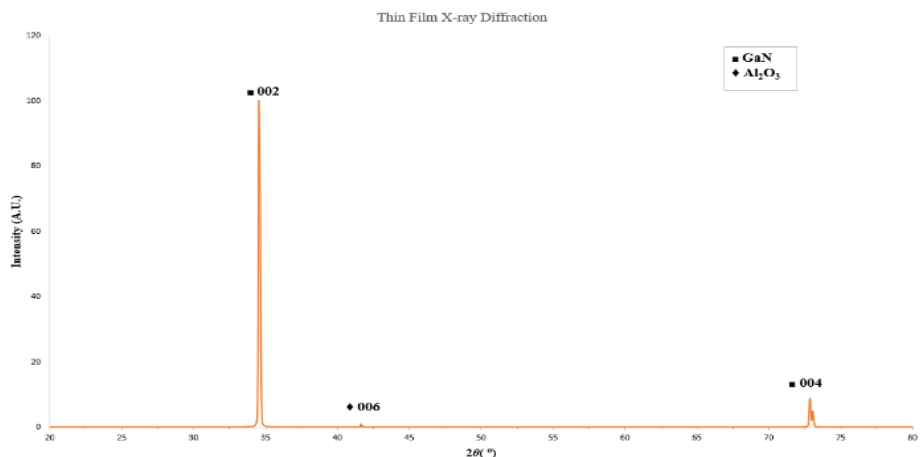

**Fig. 6.** X-ray spectrum for thin film sample

**Table 4.** Thin film crystal planes comparison with the ICDD data

| Experimental 2θ (°) | Real Intensity | Relative Intensity | ICDD 2θ(°) | ICDD Relative Intensity | 2θ Difference | d (Å)=λ/2sinθ | Material | hkl plane |
|---|---|---|---|---|---|---|---|---|
| 34.51 | 6138711 | 100 | 34.59 | 45 | 0.08 | 2.59 | GaN | 002 |
| 41.61 | 58359 | 0.95 | 41.71 | 5 | 0.1 | 2.16 | $Al_2O_3$ | 006 |
| 72.85 | 542345 | 8.83 | 72.98 | 3 | 0.13 | 1.29 | GaN | 004 |

**Table 5.** Smallest crystallite size of thin film sample

| Material | hkl plane | Relative Intensity | θ (°) | B (Rad)×$10^{-3}$ | t (nm) |
|---|---|---|---|---|---|
| GaN | 002 | 100 | 17.25 | 3.14 | 46.23 |
| $Al_2O_3$ | 006 | 0.95 | 20.81 | 4.89 | 30.33 |

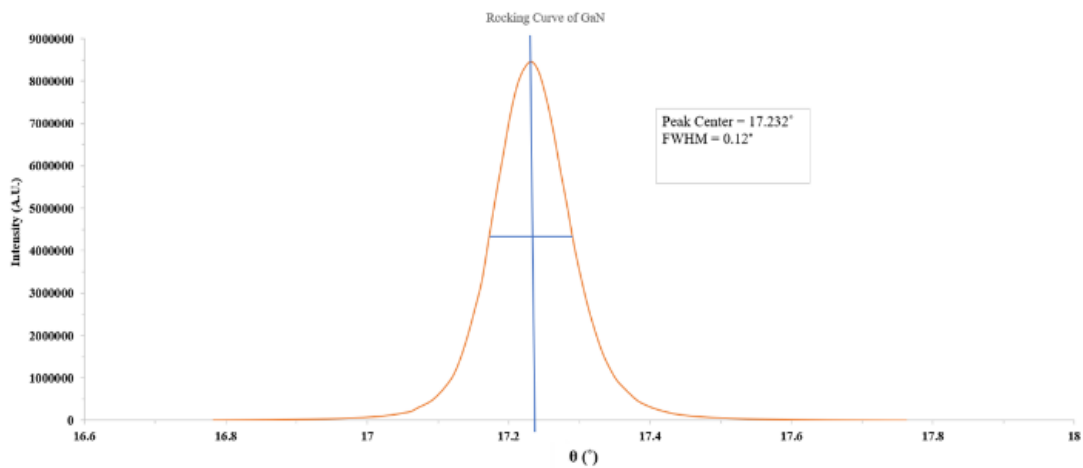

**Fig. 7.** Rocking curve of GaN

**Table 6.** Lattice Parameters for single crystal sample

| Material | Formula | a(Å) | b(Å) | c(Å) |
|---|---|---|---|---|
| GaN | $\frac{1}{d^2} = \frac{4}{3}\left(\frac{h^2+hk+l^2}{a^2}\right) + \frac{l^2}{c^2}$ | Unknown | Unknown | 5.18 |
| $Al_2O_3$ | $\frac{1}{d^2} = \frac{1}{a^2}\frac{(1+cos\alpha)\{(h^2+k^2+l^2)-(1-tan^2\frac{1}{2}\alpha)(hk+kl+lh)\}c}{1+cos\alpha-2cos^2\alpha}$ | Unknown | Unknown | Unknown |

## 4. CONCLUSIONS

In summary, this work represents a crucial step forward in the field of materials science and instrumentation by meticulously investigating the sources of measurement error in X-ray diffraction experiments and implementing effective calibration procedures. For this purpose, 2 X-ray diffraction analyses were conducted on two different samples of powder and single crystal materials, namely: a powder comprised of $BaSO_4$, ZnO and Al, as the first sample, and a single crystal sample comprised of GaN and $Al_2O_3$. Considering the results, indexing all the peak positions in the θ-2θ scan was carried out, and the important parameters like d-spacing of planes, smallest crystallite sizes, crystal structures and lattice parameters were also determined. The rocking curve of the single crystal film determined the out-of-plane alignment and FWHM value. By ensuring the accuracy and reliability of X-ray diffraction data, this work underpins the foundation of scientific research in materials characterization and instrumentation, ultimately advancing our knowledge and contributing to the robustness of scientific outcomes.